\documentclass[adraft]{eptcs}

\title{Parallel needed reduction for pure interaction nets}
\author{Anton Salikhmetov
\email{anton.salikhmetov@gmail.com}}
\def\titlerunning{Parallel needed reduction for pure interaction nets}
\def\authorrunning{A. Salikhmetov}

\usepackage[utf8]{inputenc}
\usepackage[english]{babel}
\usepackage[title]{appendix}
\usepackage{amsmath}
\usepackage{amssymb}
\usepackage{amsthm}
\usepackage{fixmath}
\usepackage{wrapfig}
\usepackage{caption}
\usepackage{subcaption}
\usepackage{tikz}
\usepackage{tikz-inet}

\DeclareFontFamily{U}{mathb}{\hyphenchar\font45}
\DeclareFontShape{U}{mathb}{m}{n}{
<-6> mathb5
<6-7> mathb6
<7-8> mathb7
<8-9> mathb8
<9-10> mathb9
<10-12> mathb10
<12-> mathb12
}{}
\DeclareSymbolFont{mathb}{U}{mathb}{m}{n}
\DeclareMathSymbol{\righttoleftarrow}{\mathrel}{mathb}{"FD}

\tikzset{every node/.style = {node distance=0em, scale=0.8}}

\newcommand{\ar}{\text{\textsf{ar}}}
\newcommand{\hole}{{[\phantom M]}}

\begin{document}
\maketitle

\begin{abstract}
Reducing interaction nets without any specific strategy benefits from constant time per step.
On the other hand, a canonical reduction step for weak reduction to interface normal form is linear by depth of terms.
In this paper, we refine the weak interaction calculus to reveal the actual cost of its reduction.
As a result, we obtain a notion of needed reduction that can be implemented in constant time per step without allowing any free ports and without sacrificing parallelism. \end{abstract}

\section{Introduction}

Previously, we successfully adapted the approach of token-passing nets~\cite{sinot} to optimal reduction~\cite{termgraph} as well as closed reduction\footnote{
The project called Macro Lambda Calculus is available at \url{https://www.npmjs.com/package/@alexo/lambda} as a Node.js package and implements the pure untyped lambda calculus using interaction nets, providing both CLI and API.
The package includes several encodings of the lambda calculus, each of them making use of the embedded read-back mechanism.
}.
However, dissatisfied with difficulties of adapting the approach of token-passing and having to leave the pure formalism of interaction nets by introducing the non-deterministic extension, we decided to consider implementation of the weak reduction to interface normal form~\cite{pinto}.

Switching to weak reduction comes at a cost.
First of all, a reduction step is no longer constant by time, but at least linear by depth of terms.
Second, weak reduction requires the notion of interface in its original interaction calculus variant, while we do not allow any free ports in our implementation of interaction nets\footnote{
The JavaScript Engine for Interaction Nets is available at \url{https://www.npmjs.com/package/inet-lib} as a Node.js package that implements a programming language based on the interaction calculus.
}.
Moreover, we avoid any notion of a root of a net in order to preserve the option of implementing a distributed computation framework based entirely on interaction nets.

These considerations have led us to refining the interaction calculus for weak reduction.
The resulted version of interaction calculus is presented in this paper.
We define reduction that can be implemented in constant time per step, thus revealing the actual cost of weak reduction.
Also, our interaction calculus formalizes the notion of needed reduction of interaction nets without allowing free ports.
And finally, the option of parallel evaluation has been preserved.

\section{Definitions}

A \textit{term} is inductively defined as ${t ::= {!}\alpha(t_1,\dots,t_n)\ |\ \alpha(t_1,\dots,t_n)\ |\ x}$, where $x$ is called a \textit{name}, $\alpha$ is an agent of type ${\alpha \in \Sigma}$ from a set $\Sigma$ that is called \textit{signature}, and ${n = \ar(\alpha) \geq 0}$ is the agent's \textit{arity}.
If a term $t$ has the form of ${!}\alpha(t_1,\dots,t_n)$, then we call $t$ \textit{needed} and denote it as ${!}t$.
An \textit{interaction rule} is an unordered pair ${\alpha[v_1,\dots, v_m] \bowtie \beta[w_1,\dots, w_n]}$, where $m = \ar(\alpha)$, $n = \ar(\beta)$, and $v_i$ and $w_i$ are terms.
Signature and a set of interaction rules together define an \textit{interaction system}.
In any interaction system, a \textit{configuration} is defined as an unordered multiset of \textit{equations} $v_i = w_i$ denoted as ${\langle v_1 = w_1,\dots,v_n = w_n\rangle}$.
Any name $x$ can have either zero, or exactly two occurrences in a configuration.
If a name $x$ has exactly one occurrence in a term $t$, then 
\textit{substitution} $t[x := u]$ is the result of replacing $x$ in $t$ with the term $u$.

\clearpage
\section{Reduction}

Reduction relation on configurations is defined for three different cases.

If $\alpha[v_1,\dots, v_m] \bowtie \beta[w_1,\dots, w_n]$, then the following reduction is called \textit{interaction}:
$$
\langle {!}\alpha(t_1,\dots,t_m) = {(!)}\beta(u_1,\dots,u_n),\ \Delta \rangle
\rightarrow
\langle t_1 = v_1,\dots,\ t_m = v_m,\ u_1 = w_1,\dots,\ u_n = w_n,\ \Delta \rangle,
$$
where $(!)$ stands for either $!$ or absence of it.

The second case of reduction is \textit{indirection} defined for a name $x$ that occurs in $v$:
$$
\langle x = t,\ v = w,\ \Delta \rangle
\rightarrow
\langle v[x := t] = w,\ \Delta \rangle
.
$$

Finally, the following reduction is called \textit{delegation}:
$$
\langle v[x := \alpha(t_1,\dots, {!}t_i,\dots,t_n)] = w,\ \Delta \rangle
\rightarrow
\langle v[x := !\alpha(t_1,\dots, {!}t_i,\dots,t_n)] = w,\ \Delta \rangle,
$$
meaning that a needed term makes its parent agent needed as well.

Interaction, indirection, and delegation together constitute the reduction relation of configurations.

\section{Example}

The following reduction sequence corresponds to read-back of $\omega \equiv \lambda x.x\ x$ as defined in \cite{termgraph}:
\begin{align*}
\langle r_\hole(!p) &= \lambda(\delta(x,\ @(x,\ y)),\ y) \rangle \rightarrow
&\text{(1 delegation)} \\
\langle !r_{\hole}(!p) &= \lambda(\delta(x,\ @(x,\ y)),\ y) \rangle \rightarrow^*
&\text{(1 interaction and 2 indirections)} \\
\langle \delta(x,\ @(x,\ r_{\lambda x.\hole}(!p))) &= a_x \rangle \rightarrow^*
&\text{(3 delegations)} \\
\langle !\delta(x,\ !@(x,\ !r_{\lambda x.\hole}(!p))) &= a_x \rangle \rightarrow^*
&\text{(1 interaction and 1 indirection)} \\
\langle !@(a_x,\ !r_{\lambda x.\hole}(!p)) &= a_x \rangle \rightarrow^*
&\text{(1 interaction and 1 indirection)} \\
\langle r_{x\ \hole}(!r_{\lambda x.\hole}(!p))) &= a_x \rangle \rightarrow
&\text{(1 delegation)} \\
\langle !r_{x\ \hole}(!r_{\lambda x.\hole}(!p))) &= a_x \rangle \rightarrow
&\text{(1 interaction)} \\
\langle !r_{\lambda x.\hole}(!p) &= a_{x\ x} \rangle \rightarrow
&\text{(1 interaction)} \\
\langle !p &= a_{\lambda x.x\ x} \rangle,
&\text{(normal form)}
\end{align*}
including $5$ delegations in addition to $5$ interactions and $4$ indirections.

\section{Implementation}

A natural implementation of full reduction for interaction nets is to have a queue of pairs of single-linked trees to represent the set of active pairs with each name in configuration represented as a pair of nodes linked to each other.
That immediately gives a constant time per reduction step which is either interaction or indirection.
Note that such a queue can be processed in any order and even in parallel.

However, in case of the weak reduction, that data structure is not enough.
In addition to links from a parent node to each of its children, one could choose to add backward links from each node to its parent node, the outermost node's parent link pointing to the equation in which the corresponding term occurs.

Now, let us discuss how to implement the refined interaction calculus we introduced in this paper, aiming to preserve the good properties of the queue as noted above.
First of all, instead of the queue of active pairs, we can choose to have a queue of needed entities which can be needed terms or equations.
Processing such a queue is essentially replacing each needed node in the queue with its parent node and marking it needed as well.
When a node's parent link points to an equation, the node is to be replaced with that equation in the queue.
Implementation of interaction and indirection remains the same with the following two modifications.
First, only needed equations are to be added to the queue after interaction.
Second, after substitution of a needed term, that term is to be added to the queue.

\section{Conclusion}

Here, we introduced a version of interaction calculus that captures the notion of parallel needed reduction for pure interaction nets.
Then, we discussed one possible way to implement it in software.
The refined interaction calculus benefits from a constant time per step, reveals the actual cost of weak reduction, and preserves the option of parallel evaluation of interaction nets.
Further, we would like to study its properties more carefully and compare its implementation with others.
We expect some performance gain compared to the approach of token-passing, since implementation of delegation can be an order of magnitude cheaper than that of interaction.

\nocite{*}
\bibliographystyle{eptcs}
\bibliography{cite}
\end{document}